\documentclass[twocolumn,aps,pra,floatfix,showpacs,noshowkeys,epsfig,graphics]{revtex4}%
\usepackage{graphicx}
\usepackage{amsmath}
\usepackage{amsfonts}
\usepackage{amssymb}

\newcommand{\newc}{\newcommand}
\newc{\beq}    {\begin{equation}}
\newc{\eeq}    {\end{equation}}

\newc{\beqa}    {\begin{eqnarray}}
\newc{\eeqa}    {\end{eqnarray}}
\newc{\bs}    {\section}
\newc{\no}    {\\ \nonumber}

\newc{\st}    {\stackrel}

\begin{document}
\title{BEC dark matter can explain collisions of galaxy clusters}
\author{Jae-Weon Lee}\email{scikid@kias.re.kr}
\affiliation{School of Computational Sciences,
             Korea Institute for Advanced Study,
             207-43 Cheongnyangni 2-dong, Dongdaemun-gu, Seoul 130-012, Korea}
             \author{Sooil Lim}
 \affiliation{Department of Physics and Astronomy, Seoul National University, Seoul, 151-747, Korea}

\author{Dale Choi}\email{scikid@kias.re.kr}
    \affiliation{
Korea Institute of Science and Technology Information, Kwahanro 335, Yuseong-Gu, Daejeon, 305-806, South Korea
}

\date{\today}
\begin{abstract}
We suggest that
the dark matter model based on Bose Einstein condensate or scalar field can
resolve the apparently  contradictory behaviors of dark matter
in  the Abell 520 and the Bullet cluster.
 During a collision of two galaxies in the cluster,
if initial kinetic energy of the galaxies is large enough, two dark matter halos pass each other
in a soliton-like way as observed in the Bullet cluster. If not, the halos merge
due to the tiny repulsive interaction among dark matter particles as observed in the Abell 520.
 This idea can also explain  the origin of the dark galaxy and the galaxy without dark matter.
\end{abstract}
\maketitle

Dark matter (DM)
 constituting about 24 percent of the mass of the universe
 is one of the big puzzles in modern physics and cosmology~\cite{DMreview,dark} .
According to  numerical simulations, while the cold dark matter (CDM) with the cosmological constant  model (i.e., $\Lambda$CDM)
  is remarkably successful in explaining the formation of the structure  larger than galaxies, it seems to
encounter problems on the scale of  galaxy or sub-galactic
structure. $\Lambda$CDM model usually predict the cusped central density and
too many sub-halos, and too small angular momentum of the galaxies,
  which are, arguably, in contradiction with observations~\cite{navarro-1996-462,deblok-2002,crisis}.
   On the other hand the models based on Bose Einstein condensate (BEC) DM
  or scalar field dark matter (SFDM)  of ultra-light scalar particles well explain the observed rotation
   curves of galaxies ~\cite{PhysRevD.68.023511,Boehmer:2007um} and solve the above problems of  CDM models
     ~\cite{corePeebles,Riotto:2000kh,PhysRevD.62.103517,0264-9381-17-13-101,PhysRevD.63.063506}.

 The mystery deepened further after recent observations of  massive intergalactic collisions
 in two clusters of galaxy; the bullet cluster (1E0657-56) ~\cite{bullet}
 and the Abell 520~\cite{abell520}.
Galaxy clusters are composed of three main components behaving differently during collision; galaxies composed of stars, hot gas
 between the galaxies, and DM~\cite{cluster}.
 According to the prevailing theories,  DM composed of very weakly interacting particles
 moves only under the influence of gravity and is presumed to be collisionless.
Since the stars are sparse, they can be also treated as effectively
 collisionless particles.
Thus, when two clusters of galaxy collide, we expect stars and dark matter to move together
 even during a violent
 collision,
 while intergalactic gases self-interact   electromagnetically
  and lag behind the other matters at the collision center.
  The distribution  of
DM can be inferred by
 optical telescopes using the gravitational lensing effect, while that of the hot gases
 by X-ray telescopes like Chandra.
 The observation of the Bullet cluster ~\cite{bullet} using these telescopes
 seems to be consistent with this expectation,
 and to support the collisionless CDM theory.
On the contrary, in the Abell 520,
galaxies (stars) were
  stripped away from the central dense core of the cluster,
where gases and DM are left.
This indicates DM as well as gases is collisional, which is puzzling.
The collision separating DM from visible matter is
also recently observed~\cite{ring} in the ring-like structure in the galaxy cluster Cl 0024+17~\cite{Jee:2007nx}.
It is very hard to explain the contradictive behaviors of DM in these clusters in the context of
the standard CDM model or even with the modified gravity theories.

In this paper, we suggest that this contradiction can be also readily resolved in
BEC/SFDM model. Furthermore, our theory can explain the origin of the
dark galaxy and the galaxy without dark matter.

First, let us  briefly review BEC/SFDM model.
In 1992, to explain the observed  galactic rotation curves, Sin \cite{sin1,sin2} suggested that galactic halos
are   astronomical objects in BEC  of
ultra light DM particles such as
pseudo Nambu-Goldstone boson (PNGB)
which  have Compton wavelength
$\lambda_{comp} = {h}/{mc} \sim 10~pc$, i.e. $m \simeq 10^{-24} eV$. In this model the halos are like gigantic atoms where
cold boson DM particles are condensated in a single macroscopic wave function
 and the quantum mechanical uncertainty principle
provides a force against
 self-gravitational collapse.
In the same year one of the author (Lee) and Koh ~\cite{kps,myhalo} generalized Sin's BEC model by considering a repulsive self-interaction
among DM particles, in the context of field theory and the general relativity.
(See \cite{myreview} for a review.)
In this model a BEC DM halo is a giant boson star (boson halo~\cite{review,review2})
surrounding a visible matter of galaxy and is  described by
 a coherent complex scalar field $\phi$ having a typical action
\beq
\label{action}
 S=\int \sqrt{-g} d^4x[\frac{-R}{16\pi G}
-\frac{g^{\mu\nu}} {2} \phi^*_{;\mu}\phi_{;\nu}
 -U(\phi)]
\eeq
with a repulsive potential
$U(\phi)=\frac{m^2}{2}|\phi|^2+\frac{\lambda}{4}|\phi|^4$.
 It was found that~\cite{myhalo} there are constraints
$\lambda^{\frac{1}{2}}(\frac{M_P}{m})^2\st{>}{\sim} 10^{50}$,
and $10^{-24}~eV\st{<}{\sim} m \st{<}{\sim}10^3~eV$, where $M_P$ is the Planck mass.
 We will call two models
 as BEC/SFDM model ~\cite{guzman:024023}.

In this model, for $\lambda=0$, the formation of DM structures smaller than the Compton wavelength
 is suppressed by the uncertainty principle
and this property could alleviate the aforementioned problems of the $\Lambda$CDM model.
For  $\lambda\neq0$ the minimum scale becomes $\Lambda^{1/2}/m$,
where a dimensionless coupling term
 $\Lambda= \lambda M_P^2/4\pi m^2$
is very large even for very small $\lambda$
 due to the smallness of $m$
 relative to $M_P$. Thus, the
self-interaction effect is non-negligible, if $\lambda\neq0$.
Despite of their tiny mass, BEC DM particles act as CDM particles~\cite{Matos:2003pe}
for the cosmological structure formation, because
their velocity dispersion is very small.
Thus, BEC/SFDM is an ideal
alternative to the standard CDM
 playing a role of CDM at the scale larger than a galaxy,
and at the same time suppressing sub-galactic structures.
Later  similar ideas were  rediscovered  by many authors~\cite {Schunck:1998nq,PhysRevLett.84.3037,PhysRevD.64.123528,PhysRevD.65.083514,
repulsive,Fuzzy,corePeebles,Nontopological,PhysRevD.62.103517,Alcubierre:2001ea,
Fuchs:2004xe,Matos:2001ps,0264-9381-18-17-101,PhysRevD.63.125016,Julien,moffat-2006}.
(See ~\cite{DMmodels} for a  review.)

Now, we investigate in detail the idea that
the repulsive self-interaction between BEC DM particles separate
stars from DM during the collision between galaxies or cluster of galaxies.
To do this we need equations describing the motion of DM halos.
\begin{figure}[ht]
\includegraphics[scale=.5]{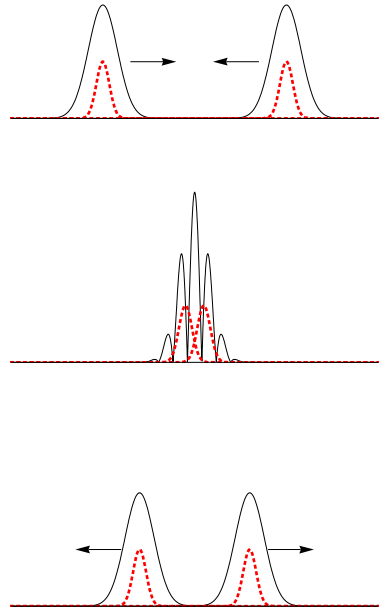}
\caption{(Color online) Schematic
diagram representing
axisymmetric collision of two galaxies. The black curves represent the distribution of BEC DM in galactic halos
and the red dotted lines that of stars of the galaxies.
If the initial kinetic energy is high enough,  DM and the stars in each galaxy
 move together
after the collision like solitons.
This could be what happened to galaxies in the Bullet cluster.}
\label{fig:figure1}
\end{figure}

\begin{figure}[ht]
\includegraphics[scale=.5]{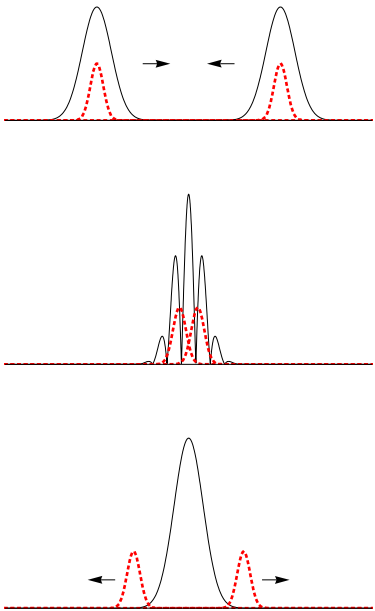}
\caption{(Color online)
The same diagram in Fig. 1 with smaller initial kinetic energy.
In this case the repulsion between DM particles
plays a significant role.
Two DM halos merge
 to form a larger DM halo,
 which can be identified as a dark galaxy,
  while stars keep going outward and could form galaxies
  without DM later.
 This could be what happened to galaxies in the Abell 520.}
\label{fig:figure2}
\end{figure}
With a spherical symmetric metric
$ ds^2=-B(r)dt^2+A(r)dr^2+r^2 d\Omega_3$
the equation of motion for the scalar field becomes~\cite{colpi}
\beq
\sigma''+[\frac{2}{x}+\frac{B'}{2B}-\frac{A'}{2A}]\sigma'
+A[(\frac{\Omega^2}{B}-1)\sigma -\Lambda \sigma^3]=0,
\label{eqs}
\eeq
where $x=mr, \Omega=\frac{\omega}{m}$ and $
\phi(x,t)=(4\pi G)^{-\frac{1}{2}} \sigma(r)
e^{-i\omega t}$.
Since the collision velocity is non-relativistic ($v\sim 10^{-3} c$)
we can use a Newtonian approximation, in which
 Einstein equation for this system reduces to
\beq
\label{poisson}
 \nabla ^2 V = 4\pi GT_{00},
\eeq
where the energy momentum tensor is
\beq
 T_{\mu \nu }  = \frac{1}{2}\left( {\partial _\mu  \phi ^* \partial _\nu  \phi  + \partial _\mu  \phi \partial _\nu  \phi ^* } \right) - g_{\mu \nu } \left( {\frac{1}{2}\partial ^\mu  \phi ^* \partial _\mu  \phi  + U(\phi )} \right)
\eeq
and  $V$ is the Newtonian gravitational potential.
 The Newtonian limit of  Eq. (\ref{eqs}) and the dimensionless form of
  Eq. (\ref{poisson})
 can be simply written as
 \beq
 \left\{ \begin{array}{l}
 \nabla _{} ^2 V = \sigma ^2  + \frac{{\Lambda \sigma ^4 }}{4} \\
  \nabla ^2 \sigma  = 2V\sigma    \\
\end{array} \right. .  \\
\label{all}
 \eeq
For $\Lambda=0$ these equations are equivalent to the non-linear Schr\"oedinger equation of
  Sin's model~\cite{newt1}.

  Since galaxy clusters are composed of about $50\sim 1000$ individual
galaxies each surrounded by galactic DM halos,
we can treat the collision of galaxy clusters as massive collisions
of individual galaxies
and expect our analysis below on the collision of two galaxies  can be applied to
the collision of clusters too.
Since DM is the major component of a galaxy, one can assume
that the collision dynamics of two galaxies is mainly governed by that of DM,
and baryonic matter (stars and gases) plays a passive role during the collision.

Choi and others ~\cite{choi1,choi2,collision3,collision1} numerically studied the head-on collision of the
boson stars described by Eq. (\ref{all}).
It was shown that  there are two regimes with very
different dynamical properties: solitonic and merging regimes.
If two colliding boson stars (galactic DM halos in
our theory)
have large enough kinetic energy,
then the halos pass each other like solitons during the collision.
We argue that this is just what happened to galactic DM halos  in the Bullet cluster.
In this case the total energy $E$ which is composed of kinetic energy $K$,
the gravitational potential energy $W$ and the repulsion energy $I$ between DM particles (determined
by the term $\lambda |\phi|^4$)
should be positive, i.e., $E=K+W+I>0$ ~\cite{choi2,collision3,collision1}.
In other words, initial relative velocity of colliding galaxies should be large enough
to overcome the  self-gravitational attraction
and repulsion force between DM particles.
Fig. 1. shows the schematic
diagram representing
collision of two galaxies. (We ignored the hot intergalactic gases which mainly exist between
the galaxies. This does not change our conclusion significantly.)
If the initial kinetic energy is large enough,  DM and the stars in each galaxy
 move together
even after the collision like solitons.
This could be what happened to galaxies in the Bullet cluster.

On the contrary, if the kinetic energy is small so that
$E=K+W+I<0$, they merge
to form a single large DM halos as shown in Fig. 2.
This could be what  happened  to  DM halos in the Abell 520.
Since our model treat galactic DM halos as boson stars, two
different regimes of the boson star collision explain
the observed contradictive behaviors of DM in two clusters.
The DM halos in the Abell 520 did not have an enough velocity and was even slowed by the repulsion, while the stars,
having the same initial velocity, managed to escape the potential well because they are collisionless.
This situation can happen only when the collision velocity is within an appropriate range.
We expect usually
 the collision velocity of other colliding clusters or galaxies is too slow or fast to separate  efficiently stars from DM halos.
This explain why stars usually trace DM.
The large DM halo left at the center can be identified as a dark galaxy~\cite{darkgalaxy} like  VIRGOHI21,
  while two star groups going outward  could form two independent galaxies
  without DM later~\cite{galaxywodm}, arguably, like M94 (NGC 4736)~\cite{galaxywodm}.
  The origin of these galaxies was a mystery so far.
Thus, our theory  explain not only the mystery of galaxy clusters but also
the origin of the dark galaxy and the  galaxy without DM.
This scenario also implies that there are many dark galaxies at the center of the Abell 520
and galaxies without or very small DM at outermost region of the cluster.

For the scenario to be plausible the initial collision velocity of galaxies in
the Bullet cluster should
be much larger than that of the Abell 520.
Interestingly, according to the observations ~\cite{bullet,abell520}, the Abell 520 actually had much small collision velocity than
the Bullet cluster.
The estimated collision velocity of the clusters inferred from the X-ray temperature
of the gases are
about $4700 km/s$
and $1000 km/s$
for the Bullet cluster and the Abell 520, respectively.
This observational data support
our theory.

Our theory provides a possibility of determining the mass $m$ and self-coupling $\lambda$
of DM particles using the data from the collision of galaxy clusters.
Recently, it is also suggested that the observed size evolution of very massive galaxies
  and the early compact galaxies can be also well explained in BEC/SFDM model~\cite{Lee:2008ux}.
In conclusion,
since BEC/SFDM model have passed many tests and explain many mysteries of galaxies and galaxy clusters
which seems to be hardly possible in other DM theories,
this model can be a promising alternative to the usual CDM model.

\section*{ ACKNOWLEDGMENTS }

\vskip 5.4mm


\end{document}